\documentclass{myarticle}
\usepackage{amsmath,amsfonts,amssymb}
\usepackage{eucal}
\usepackage{mathtools}
\usepackage{accents} 
\usepackage{lineno}
\usepackage[hypertexnames=false]{hyperref} 
\usepackage{graphicx}
\usepackage{relsize}
\usepackage[nice]{nicefrac}
\usepackage{subfig}
\usepackage{comment}
\usepackage{mathrsfs}

\usepackage{float} % to force positioning a figure with the [H] option

\biboptions{sort&compress}  %% multiple citations in the form type i.e. [1–4]
\newcommand{\gend}{\mathfrak{g}}
\newcommand{\remd}{\mathfrak{r}}

\newcommand{\eqdef}{=\vcentcolon}

\newcommand{\sub}[1] {_{_{\text{\scriptsize #1}}}}
\newcommand{\subs}[1] {_{_{\text{\tiny #1}}}}

\newcommand{\msub}[1] {_{_{\text{\scriptsize $#1$}}}}

\newcommand{\suchthat}{\mathlarger{\backepsilon}\,'}

\newcommand{\bcheck}[1]{\ensuremath{\accentset{\vee}{#1}}}

\begin{document}

\begin{frontmatter}
  
  \title{The SIR model towards the data.\\
   \small{ One year of Covid-19 pandemic in Italy case study and plausible ``real'' numbers.}\\ \vspace{10pt}\scriptsize{(ACCEPTED {\em as is} at The European Physical Journal Plus on July 21, 2021)}}

\author[mysecondaryaddress]{I. Lazzizzera\corref{mycorrespondingauthor}}
\address{Associated with Department of Physics - University of Trento - Italy}
\address{Associated with  Trento Institute for Fundamental Physics and Applications - INFN - Italy}
\address{via Sommarive 14 - 38123 Povo (TN) Italy}
\cortext[mycorrespondingauthor]{Corresponding author}
\ead{ignazio.lazzizzera@unitn.it}

\begin{abstract}
In this work, the SIR epidemiological model is reformulated so to highlight the important {\em effective reproduction number}, as well as to account for the {\em generation time}, inverse of the {\em incidence rate}, and the {\em infectious period} (or {\em removal period}), inverse of the {\em removal rate}. The aim is to check whether the relationships the model poses among the various observables are actually found in the data. The study case of the second through the third wave of the Covid-19 pandemic in Italy is taken. Given its scale invariance, initially the model is tested with reference to the curve of swab-confirmed infectious individuals only. It is found to match the data if the given curve of the {\em removed} (that is healed or deceased) individuals is assumed underestimated by a factor of about 3 together with other related curves. Contextually, the {\em generation time} and the {\em removal period}, as well as the {\em effective reproduction number}, are obtained fitting the SIR equations to the data; the outcomes prove to be in good agreement with those of other works. Then, using knowledge of the proportion of Covid-19 transmissions likely occurring from individuals who didn't develop symptoms, thus mainly undetected, an estimate of the {\em ``true numbers''} of the epidemic is obtained, looking also in good agreement with results from other, completely different works. The line of this work is new and the procedures are computationally really inexpensive.

\end{abstract}

\begin{keyword}
\texttt{\\SIR epidemic model, Kermack-McKendrick model, data analysis, Covid-19 true numbers, symptomatic, asymptomatic.}
\end{keyword}

\end{frontmatter}

%\linenumbers

\section{Introduction}
The \textbf{SIR model} \cite{Kermack_McKendrick, Murray:1993, Daley_Gani, Brauer:2017, Martcheva2015a, Brauer_Castillo-Chavez_Feng}, developed by Kermack and McKendrick \cite{Kermack_McKendrick} in 1927, is the well known very simple model of infectious diseases that considers {\em three-compartments}, recalled here to state terminology and notations: \newline
the compartment \textbf{S} of susceptible individuals; \newline
the compartment \textbf{I} of the {\em infectious} (or {\em currently positive}) individuals, who have been infected and are capable of infecting susceptible individuals during the {\em infectious period}; \newline
the compartment \textbf{R} of the {\em removed} individuals, who recovered from the disease or died from the disease, the former assumed to remain immune afterwards. \newline
Births and non-epidemic-related deaths are neglected. \newline
The cardinality of each of the compartments are indicated with the corresponding non bold letters, while $N$ denotes the involved total population at an initial time $t_0$:
\begin{equation} 
  S(t_0) + I(t_0) + R(t_0) = N\,. \label{N const}
\end{equation} 
The disease {\it incidence rate} $\beta$ is defined so that $\beta\,S\,I$ gives the number of new infections per unit time \cite{Martcheva2015a}; the {\it removal rate} $\gamma$ is defined so that $\gamma\,I$ gives the rate at which infectious individuals “deactivate” (heal or die). Typically, experts adopt $\beta$ constant over time, which is not the general case, due to possible mutations of the decease carrier or social measures to counter the spread of the infection; also, to simplify mathematics, they adopt equations with null {\it generation time} $\gend$, that is the infector-infected pairing time lapse, as well as null {\it removal period} $\remd$, which is the average time between infection and recovery or death, despite the relation
\begin{equation}
   \remd \,=\, \frac{1}{\gamma}\,. \label{removal pariod/rate}
\end{equation}
Within the removal period $\remd$, a typical infectious individual is expected to cause $\remd\,\beta\,S$ new infections, so defining a function of time that, normalized, is called {\em effective reproduction number} $\mathcal{R}_t $ (see for instance\cite{Hethcote2000}); namely:
\begin{equation}
  \mathcal{R}_t \,=\, \remd\;\,\beta(t)\,\frac{S(t)}{N} \,=\,
  \frac{\beta(t)}{\gamma}\,\frac{S(t)}{N}\,. \label{reproduction number}
\end{equation}
%%%%%%%%%%%%%%
\begin{comment}
  The {\em basic} reproduction number $\mathcal{R}_0$ \cite{Heesterbeek} is also defined, as the effective reproduction number at an initial time, when one infectious subject appears among an all-susceptible population (\cite{Edelstein-Keshet},  \cite{Hethcote2000}).
\end{comment}
%%%%%%%%%%%%%
\newline
So, the SIR equations as used here become: 
\begin{subequations} \label{fond.eqs.} 
  \begin{align} 
    &\frac{dS}{dt}(t+\gend) = -\,\gamma\,\mathcal{R}_t\; I(t)\,, \label{fond.eqs.1}\\
    &\frac{dI}{dt}(t+\gend) = \gamma\,\mathcal{R}_t\; I(t)\,-\,\gamma\; I(t+\gend-\remd)\,,
    \label{fond.eqs.2}\\ 
    &\frac{dR}{dt}(t+\remd) = \gamma\,I(t)\,. \label{fond.eqs.3}   
  \end{align}
\end{subequations}
As well known, they imply that the sum \,$S+I+R$\, is conserved, so that \,$S(t)+I(t)+R(t) =  N$\, at any time $t$.

\section{Outlines of the work}
Purpose of this work is to check whether the relations established in system \ref{fond.eqs.} are actually found in the data or, at least, whether ``corrections'', accounting for incomplete data or systematic errors, may or should be introduced, with the implication that consequently the relationships are satisfied. Crucial is the fact that the model is scale-invariant, thus allowing to conveniently choose as a reference one sub-compartmental curve whose real data can be considered reliable, such as the swab-confirmed infectious individuals. This choice is done indeed here: swab-confirmed infectious are mostly individuals who have developed symptoms and are actually found to cover a nearly constant fraction of all the infectious people, given the circumstances that symptomatic and a-symptomatic individuals roughly are respectively fractions of the age groups of over sixty and younger people (\cite{HiddenChall}\,, \cite{age1}\,, \cite{age2}\,, and \cite{age3}). \newline
The case of the second through the third pandemic wave of Covid-19 in Italy is studied.
First, it will be shown that the relation established through eq.~\ref{fond.eqs.3} holds true if $R(t)$ is scaled by a factor that is obtained together with the {\it removal period} $\remd$\, by a least-square procedure of matching over data. Arguments will be given for how a scaling-up could be due indeed over the official data. \newline
Once $\remd$, and thus $\gamma$, are given, the {\em effective reproduction number} is obtained through eq.~\ref{fond.eqs.2}\,, reliably, despite using the swab-confirmed infection cases only, for that equation is scale-invariant on its own.\newline
The transition to {\em ``true''} numbers is finally done, correcting the swab-confirmed infectious cases for the proportion factor of transmissions that likely occur from asymptomatic subjects. The results are compared with those obtained at the MRC Centre for Global Infectious Disease Analysis, Imperial College London (ICL,\, \cite{IC}), where a completely independent approach is used.

\section{The data set}
The data set is from Italy's Department of Protezione Civile \cite{DPCdata}, lasting from 1 June 2020 to 31 May 2021. Since every weekend there was a postponement in cases recording to a few days later, according to common practice the data is smoothed via a multi-day moving average; the choice is 9 days, to systematically include a couple of days after each weekend.

\section{Swab-confirmed infectious towards daily removed curves}
\label{h2h}
\begin{figure}[ht]
  \centering
  \includegraphics[width=0.90\textwidth]{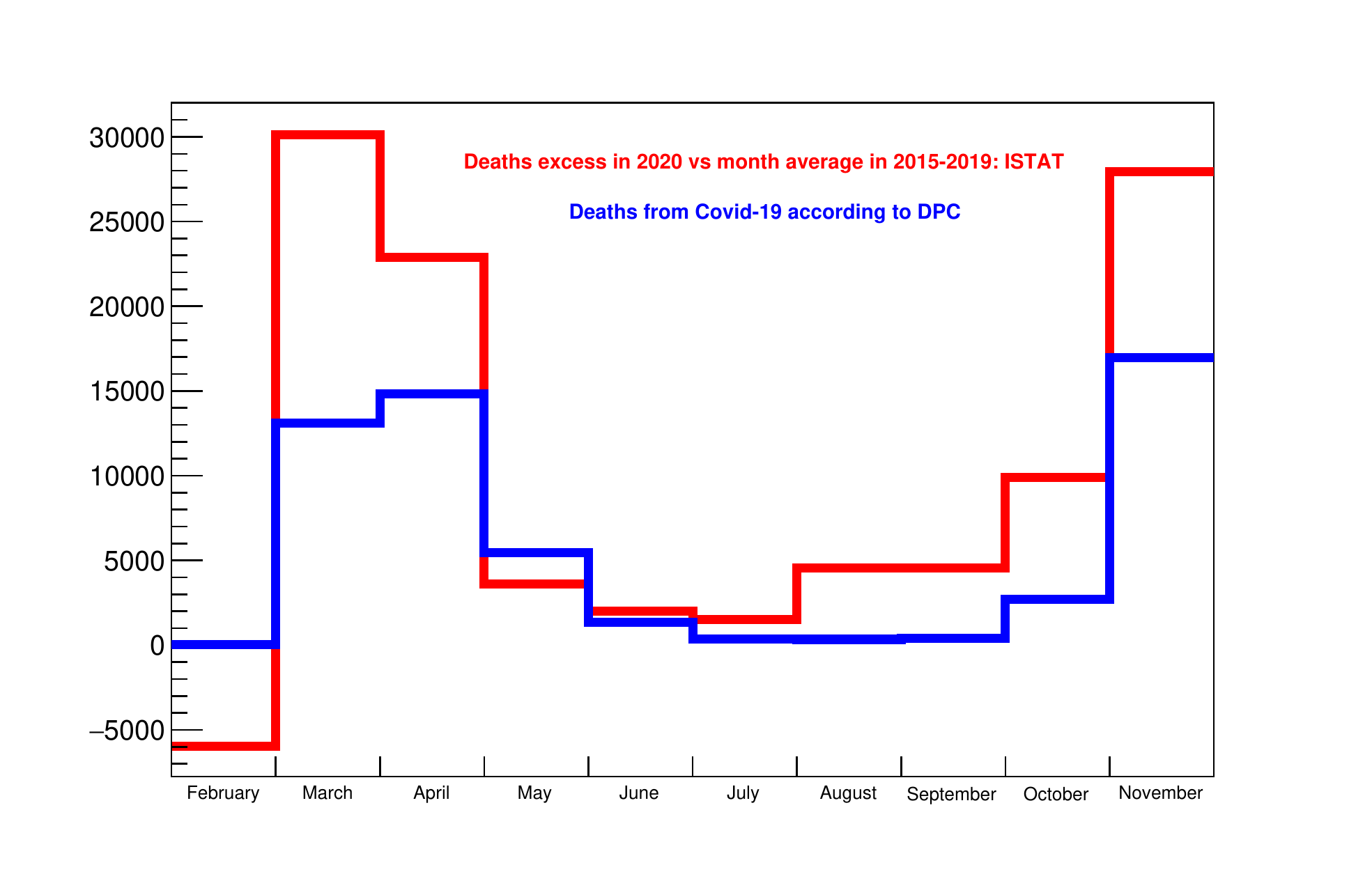}
  \caption{\footnotesize Monthly deaths from Covid-19 in Italy during 2020, according Italy's
    Department of Protezione Civile (blue histogram); monthly excess of deaths the same year
    compared to the averages over the previous five years, according Italy's ISTAT (Istituto
    Nazionale di Statistica) .}
  \label{deaths2020}
\end{figure}

Verifying that the relationship given by eq.~\ref{fond.eqs.3} is indeed found in the data is not so trivial. For example, there is evidence that the monthly deaths from Covid-19 in 2020, as given by Italy's Department of Protezione Civile, are largely underestimated: this is shown by an ISTAT study on the monthly excess of deaths in 2020, compared to the corresponding averages over the previous five years  (see \cite{ISTATdeaths2020} and \cite{BLMMOPSZ}). ISTAT is Italy's Istituto Nazionale di Statistica. The matter is illustrated in fig.~\ref{deaths2020}.
In addition to this, it is to be expected that $R(t)$ does not include most of the cases that had an asymptomatic or mild course. Also, asymptomatic infected people are probably not reported among the infectious, whereas by far most of the reported infectious are those who had swabs confirmation, whose number will be called $I_{sc}$. Let's indicate with $\bcheck{R}$ the curve of the actually registered healed and deaths: it is found that its derivative, the daily variation, shifted forward in time, is indeed proportional to $I\sub{sc}(t)$. To methodically verify eq.~\ref{fond.eqs.2}\,, the correction factor $k\subs{rel}$ is introduced so as to give maximum generality to a least-square search over the positive definite form
\begin{equation}
  \label{square form}
  \chi^2(k\subs{rel},\,\remd) \;=\;
  \sum_b\,\left[ I\sub{sc}(b)- k\subs{rel}\,\remd\,
    \frac{d\bcheck{R}}{dt}(b +\remd)\right]^2\,, 
\end{equation}
with varying $k\subs{rel}$ and the {\em removal period} $\remd$. It is worth remarking the notation $k\subs{rel}$, intended to emphasize that any correction on $\bcheck{R}(t)$, possibly {\em required} by the SIR model at this stage, is relative to the swab-confirmed infectious population only. The sum is over the days of the pandemic period considered, with the choice of weighing equally all daily data. The minimisation is performed using a C++ object of the class {\em Minimizer} of the CERN package ROOT, typically used by high energy physicist in their data analysis (\cite{ROOT}, \cite{Minimiser}): its statistical methodology is described in \cite{statsROOT}. Since the surface defined from the data through eq.~\ref{square form} is rather rough, the minimization algorithm is run 150,000 times to maximize the chance of hitting an optimal minimum:  the initial values of $k\subs{rel}$ and $\remd$ are drawn at random in the intervals $[1.0,\,5.0 ]$ and $[5.0,\,18.0 ]$ respectively. Execution on raw and smoothed data takes about one minute time altogether. The final issue for $k\subs{rel}$ and $\remd$ and their uncertainties $\delta k\subs{rel}$ and $\delta \remd$ are taken as the mean and the standard deviation of the distributions of the respective outcomes at each iterated minimization, weighted with the normalized inverse of the $\chi^2$. \newline
The results are shown on the first and the second lines of Table \ref{tab:table1}, for the raw and the smoothed data respectively. 
\begin{table}[ht!]     
  \begin{center}
    \begin{tabular}{l|c|c|c|c|c}
      % 1st column left, 2nd middle and 3rd right, with vertical lines in between
      \hline
      \,                    & $k\subs{rel}$   & $\remd$          & $\chi^2$ & ndf  & red.$\chi^2$ \\
      \hline
                             &                 &                  &          &      &             \\
      likelihood on raw data & $2.99 \pm 0.82$ & $10.31 \pm 2.74$ & $93185$  & $360$ & $266$      \\
      likelihood on smoothed & $3.14 \pm 0.82$ & $10.32 \pm 2.68$ & $32823$  & $360$ &  $91$      \\
      Gaussian fit           &                 & $ 9.78 \pm 2.28$ & $1.15$   & $32$  & $0.04$     \\
      skew sigm. derivative  &                 & $10.14 \pm 2.37$ & $2.48$   & $66$  & $0.04$     \\
      \hline
    \end{tabular}
    \caption{k-factor and $\remd$}
    \label{tab:table1}
  \end{center}
\end{table}
\newline
Since the value of the {\em removal period} $\remd$ is critical in determining $k\subs{rel}$, it is sought from the data in two further independent ways, as explained in the next two sub-sections.

\subsection{The removal period from a Gaussian fit}
At any new ``wave'' of epidemic, the rise in number of the infectious individuals follows with good
\begin{figure}[ht!]
  \centering
  \includegraphics[width=0.90\textwidth]{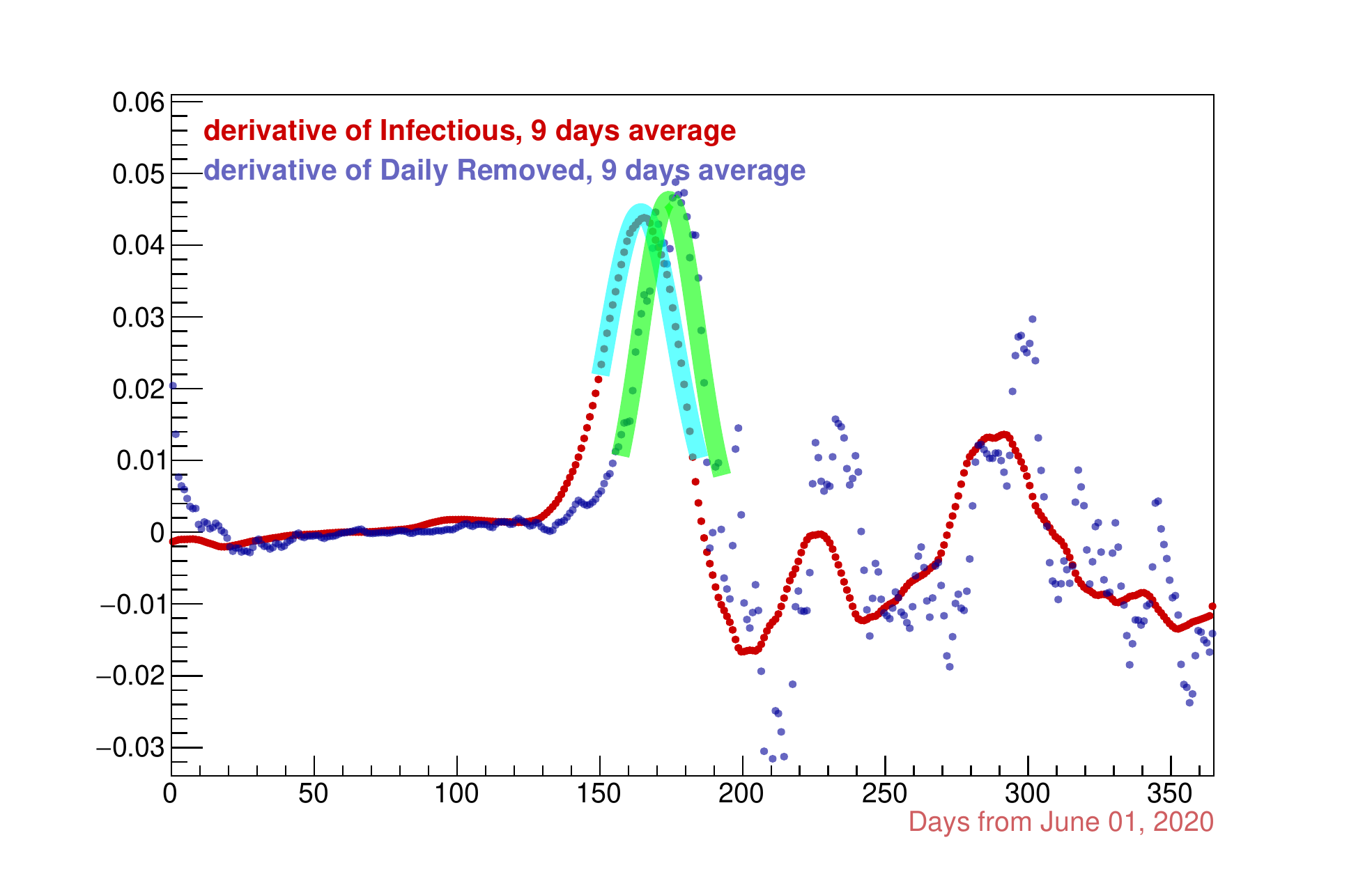}
  \caption{\footnotesize The two fitting Gaussian functions in cyan and green.}
  \label{GaussianFit}
\end{figure}
approximation a sigmoidal shape, i.e. it is roughly exponential at the very beginning, up to an inflection point, after which it bends towards a plateau; consequently, its daily variation (the time derivative) exhibits a maximum at the inflection point, around which it is approximately Gaussian.
If eq.~\ref{fond.eqs.3} correctly described the data, an analogous shape should be had in the second derivative of the {\em removal curve}. Very remarkably, this is in fact the case, as shown in fig.~\ref{GaussianFit}\,, where fitting Gaussians are plotted over the first derivative of the {\em infectious curve} and the second derivative of the {\em removal curve}: the distance in time between the vertexes of the two Gaussians gives a new independent {\em measurement} of the {\em removal period} $\remd$, reported in the third line of Table \ref{tab:table1} with associated uncertainty and fit $\chi^2$. The uncertainty is the sum in quadrature of the uncertainty on the position in time of the vertexes of the two fitting Gaussians; the $\chi^2$, with its reduced, is their worse. The fit algorithm is from the already mentioned ROOT package (CERN, \cite{TFit}).
%\vspace{25pt}

\subsection{The removal period from the ``asymmetric sigmoid derivative'' fit}
\begin{figure}[ht!]
  \centering
  \includegraphics[width=1.00\textwidth]{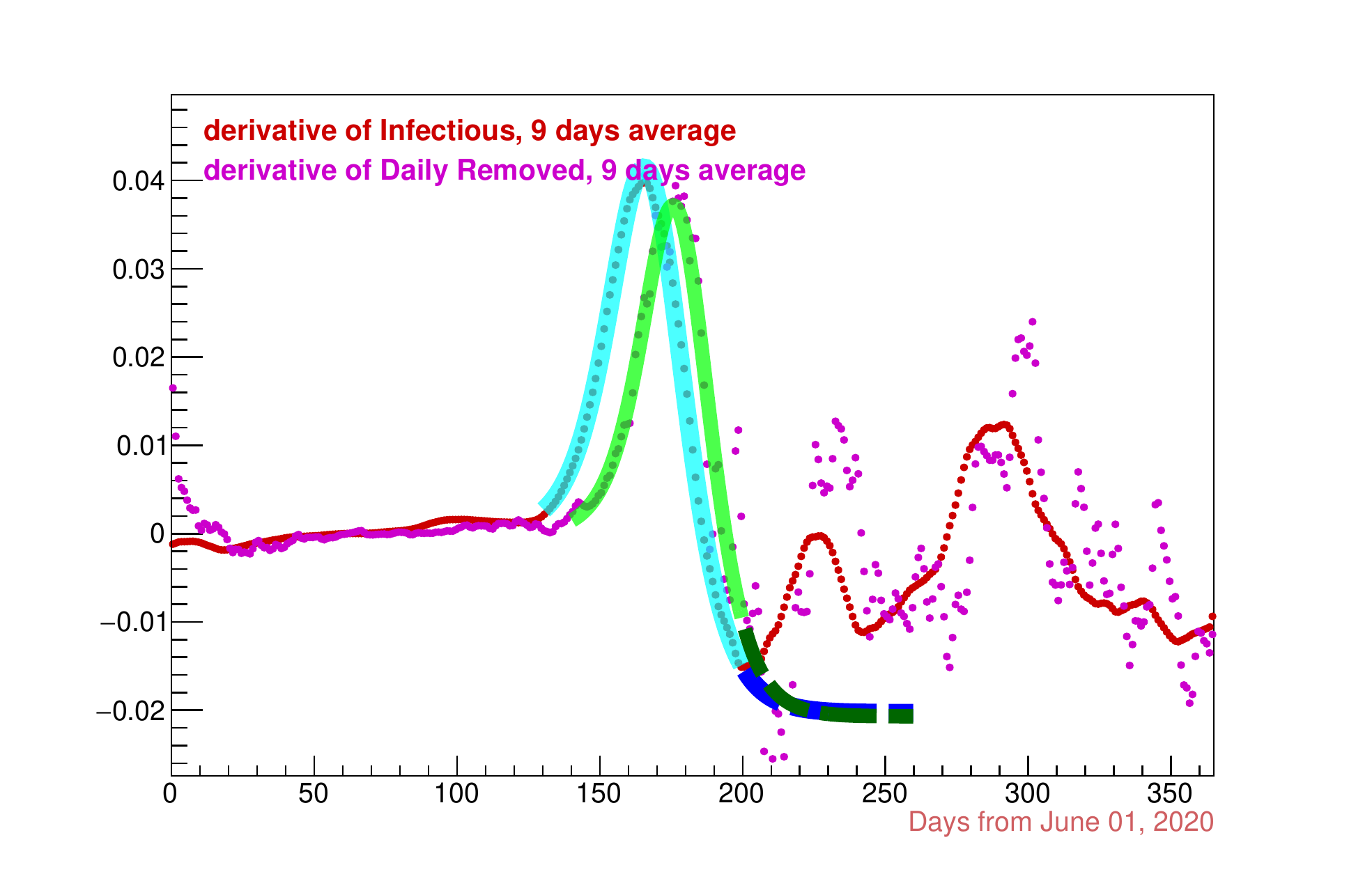}
  \caption {\footnotesize The two fitting ``asymmetric sigmoid derivative'' functions
    in cyan and green.} 
  \label{asysigder}
\end{figure}

Given the almost sigmoidal initial growth of an epidemic wave, as already mentioned in the last sub-section, an alternative fit function turns to be an asymmetric modification of the derivative of a sigmoid, which will be called {\em skew sigmoid derivative}, namely:
\begin{equation}
  \left\{
  \begin{aligned}
    s(x;\,\mu,\,\sigma,\,\epsilon) \,&=\, \frac{1}{\epsilon + (1-\epsilon)\, e^{-(x - \mu)/\sigma}}\,,
    \qquad \text{with}\quad 0 \le \epsilon < 1 \,,\\
    \mathscr{A}(x;\,\mu,\,\sigma,\,\epsilon) \, &= \,
    A\, s(x;\,\mu,\,\sigma,\,\epsilon)\, \left[ 2 - s(x;\,\mu,\,\sigma,\,\epsilon) \right] \,.
  \end{aligned}
  \right.
\end{equation}
\begin{figure}[ht!]
  \centering
  \includegraphics[width=0.60\textwidth]{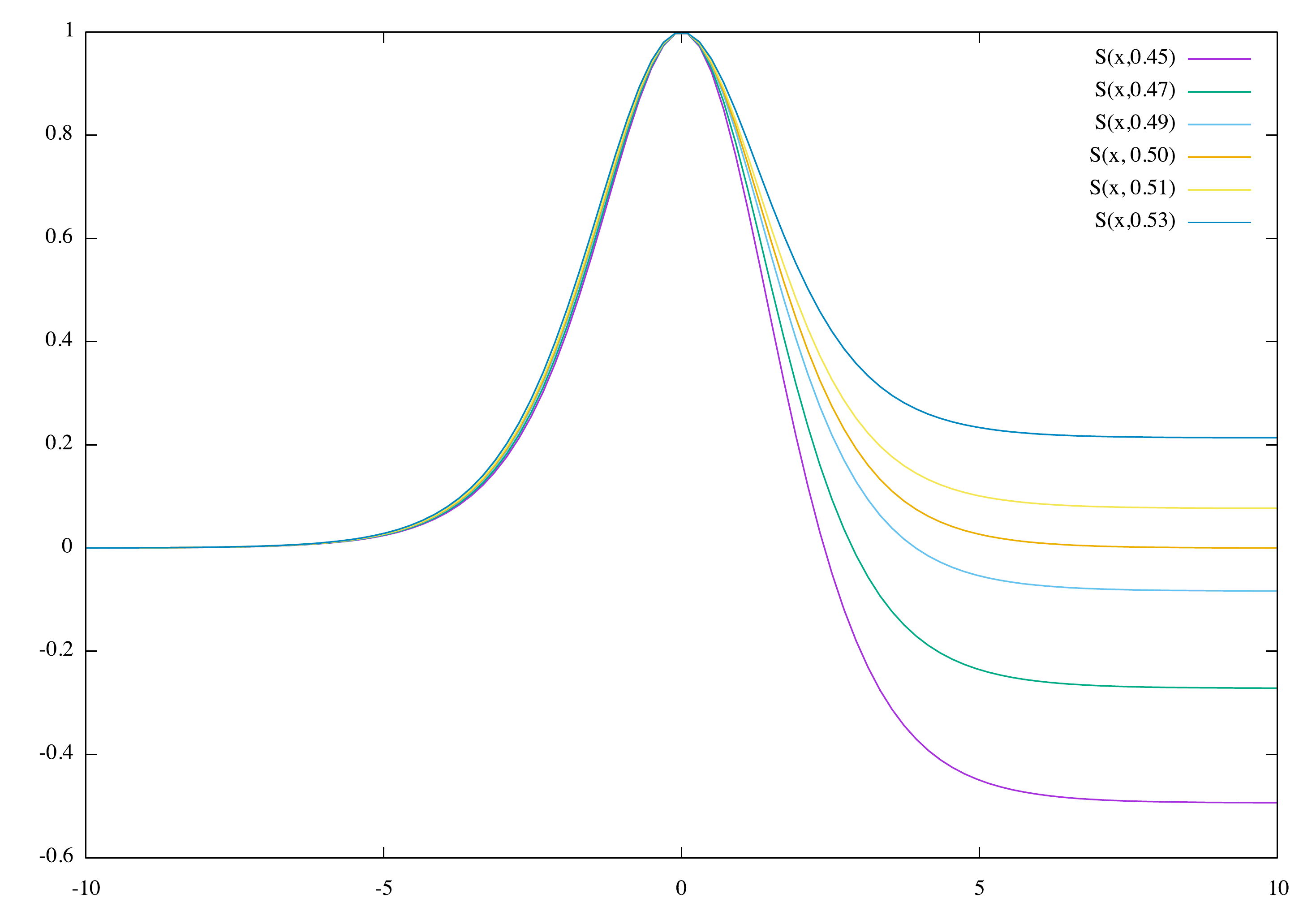}
  \caption{\footnotesize The {\em skew sigmoid derivative} for $\mu = 1.$,\, $\sigma = 1.$,
    $A = 1.$\,; $\epsilon = 0.45,\, 0.47,\, 0.49,\, 0.50,\, 0.53$.}
  \label{SigSkDer}
\end{figure}
This function has absolute maximum in $x = \mu$, with $\mathscr{A}(\mu;\,\mu,\,\sigma,\,\epsilon) = 1$. It is plotted in fig.~\ref{SigSkDer} for $\mu = 1$,\, $\sigma = 1$ and $A = 1$,
and various values of the {\em skewness} parameter $\epsilon$\,:  for $\epsilon = 0.5$ one has the derivative of a very sigmoid. The fits of the skew sigmoid derivative to the first derivative of the swab-confirmed {\em infectious curve} and to the the second derivative of the {\em removal curve} respectively are shown in fig.~\ref{asysigder}\,: again, the distance in time between the vertexes of the fitting functions gives a new {\em measurement} of the removal period $\remd$, reported in the fourth line of Table \ref{tab:table1}.

\subsection{The removal curve corrected relatively to the swab-confirmed infectious only}
\begin{figure}[ht!]
  \centering.
  \includegraphics[width=1.0\textwidth]{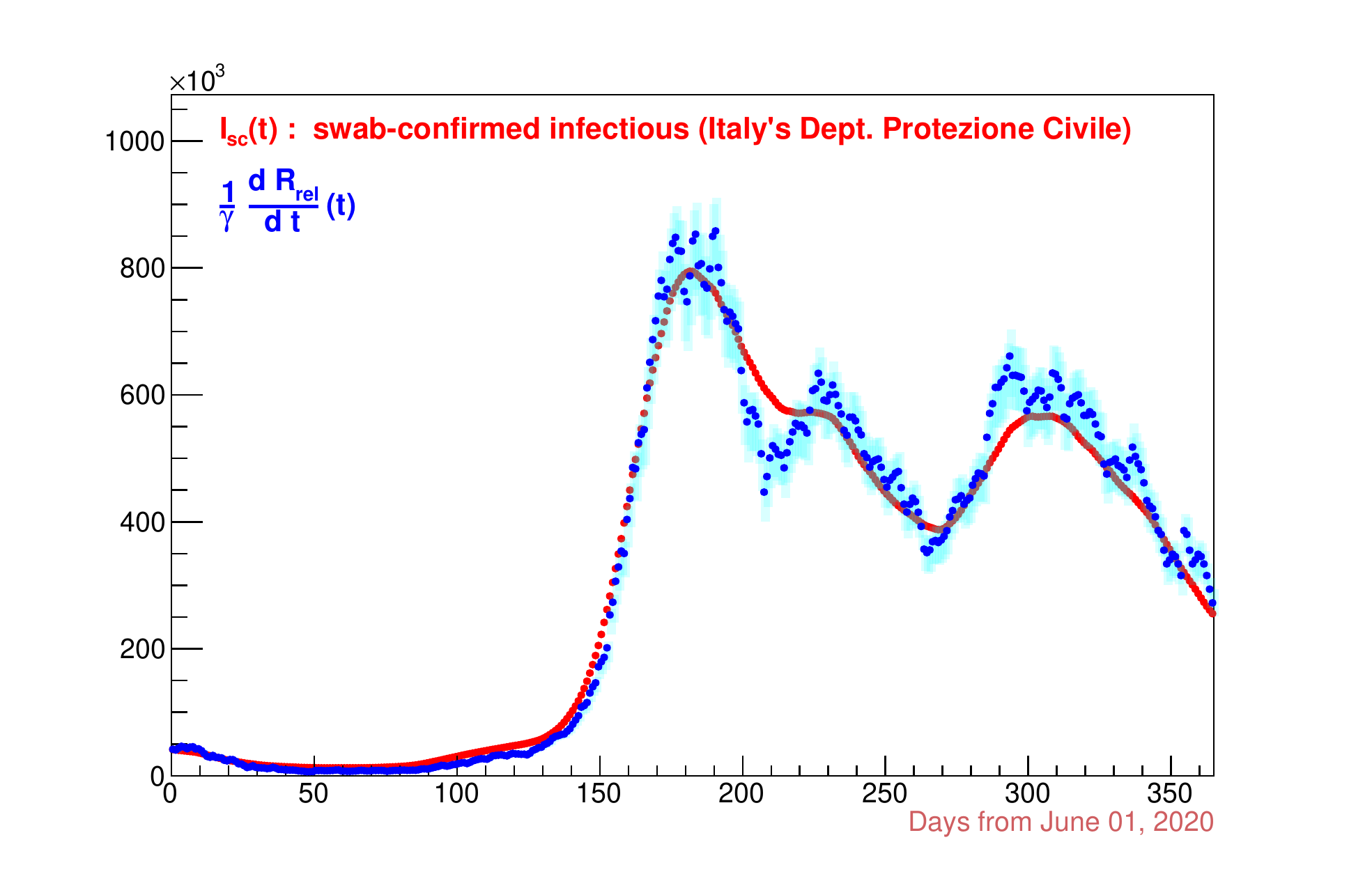}
  \caption {\footnotesize Relation \ref{rel corr}\, in the data. The derivative of
    $R\subs{rel}(t)$ (blue marks) is time-shifted by the removal period $\remd$\,.
    The cyan error bars are given by propagation from three times $\pm 0.82$ uncertainty over $k\subs{rel}$.}
  \label{scaledRemoved}
\end{figure}

From Table \ref{tab:table1} the {\em removal period} $\remd$ is assumed to be $10 \pm 2$ days, bearing in mind that the data has just one day resolution; also, comparing the $\chi^2$ on the first and second lines of the table, the correction factor $k\subs{rel}$ is taken equal to $3.14 \pm 0.82$. So we have the the curve of the {\em removed} individuals, corrected relatively to the swab-confirmed infectious only, given by:
\begin{equation}
    R\subs{rel}(t) \,=\, k\subs{rel} \, \bcheck{R}(t) \qquad \suchthat \qquad
    \frac{d R\subs{rel}}{dt}(t +\remd) \,=\, \gamma\, I\sub{sc}(t)\,. \label{rel corr}
\end{equation}
Fig.~\ref{scaledRemoved} does illustrate this: the cyan error bars are generated by the propagation of three times the $\pm 0.82$ uncertainty over $k\subs{rel}$\,.

\section{The Effective Reproduction Number}
There are several algorithms to estimate the {\em effective reproduction number} from the data: a simplified one is given in \cite{infn1}, where also an extensive bibliography on the subject can be found. The simplest yet effective estimate, that very directly interprets the meaning of the the function (see for instance \cite{Koch}), is given by 
\begin{equation}
  \mathcal{R}_t \,=\, \frac{I(t+\gend)}{I(t)}\,. \label{simple Rt}
\end{equation}
As far as the SIR model is concerned, from eq.~\ref{fond.eqs.2}\, one has
\begin{equation} \label{Rt discrete}
  \mathcal{R}_{t} \,=\,
  \frac{\remd}{I(t)}\, \frac{I(t+\gend +1)-I(t+\gend-1)}{2} \,+\,
  \frac{I(t+\gend-\remd)}{I(t)}\,.
\end{equation}
So, the derivative is implemented by the symmetric difference quotient, to have the cancellation of the first-order error in the numerical discretization. \newline
While only the {\em generation time} $\gend$ appears in eq.~\ref{simple Rt}, both $\gend$ and the {\em removal period} $\remd$ are present in eq.~\ref{Rt discrete}\,; consequently, the validity test of the SIR model through the {\em effective reproduction number} $\mathcal{R}_t$\,, it manages to provide, is to be considered quite stringent. \newline
In the previous section the {\em removal period} $\remd$ was obtained from the data using the SIR model; the question is how to get the {\em generation time} as well.
\begin{figure}[ht!]
  \centering.
  \includegraphics[width=1.0\textwidth]{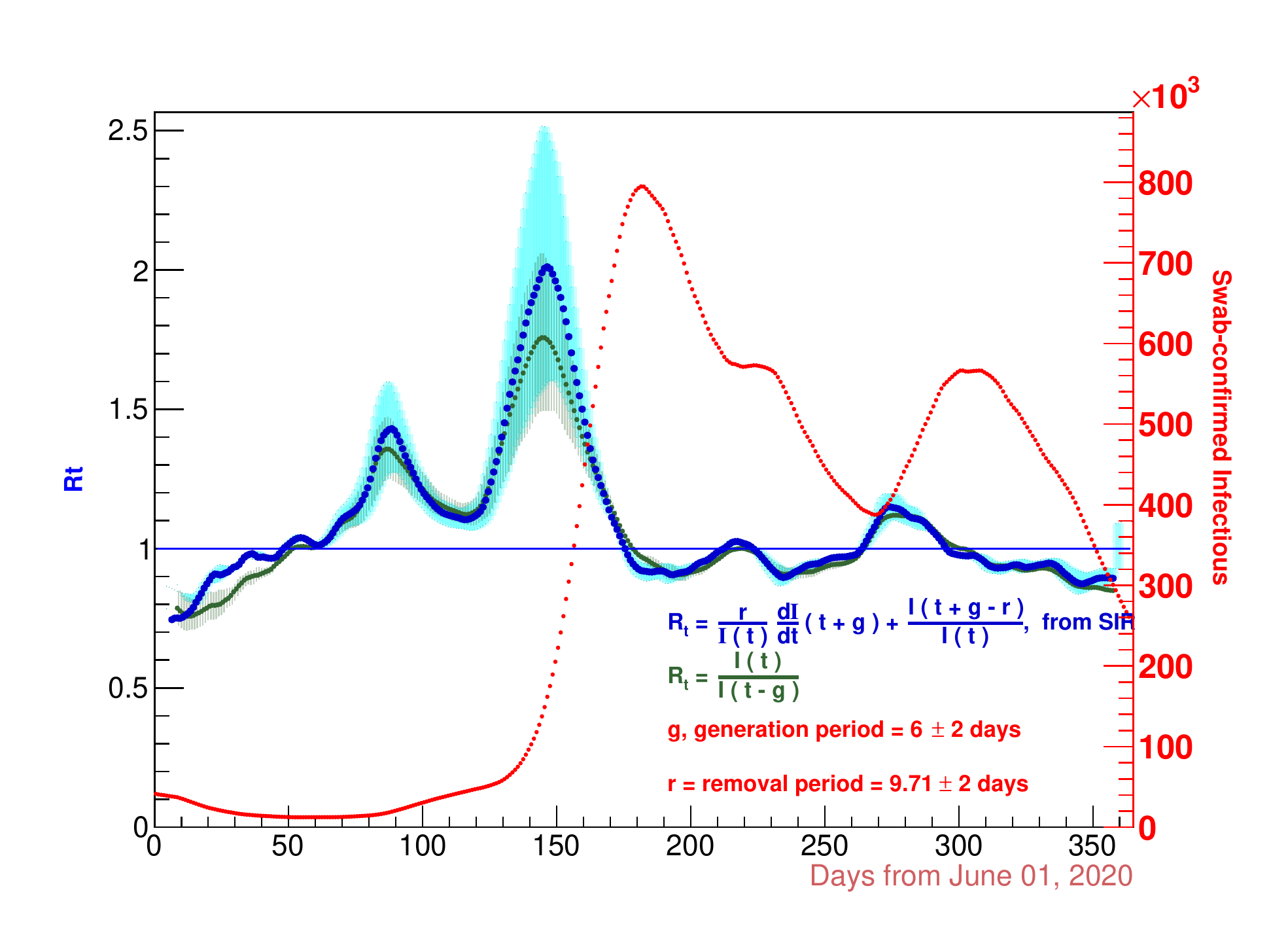}
  \caption {\footnotesize $\mathcal{R}_t$ according to the SIR model, compared with the used simplest
    formula, eq.~\ref{simple Rt}\,. In cyan are the error bars for eq.~\ref{Rt discrete}\,;
    the other are in black. $\gend = 6 \pm 2$\,; $\remd = 10 \pm\,2$.}
  \label{Fig:Rt}
\end{figure}
\newline
In our conventions, $I(t)$ denotes the total of all the infectious people, swab-confirmed or not. If $t\sub{M}$ is a day when $I(t)$ presents a maximum, then correspondingly, but $\gend$ days earlier, i.e. at day $t\sub{M} - \gend$\,, the {\em effective reproduction number} $\mathcal{R}_{t} $ should be equal to $1$, because an increase in the number of the people becoming infectious requires $\mathcal{R}_t > 1$ and a decrease requires $\mathcal{R}_t < 1$\,. Of course, every variation of $\mathcal{R}_t$ has impact on $I(t)$ with a delay of $\gend$ days, so also for $I\subs{sc}(t)$, assuming this to be proportional to $I(t)$\,. With $\remd$ fixed at $9.71 \pm 2$ days, as set out in the previous section, let's say $t\msub{\gend}$ a day when, for any given choice of $\gend$,  $\mathcal{R}_t$ is equal to $1$: in general, checking over the data, it doesn't happen that the nearest next day $t\sub{M}$, on which $I\subs{sc}(t)$ has a maximum, is such that $t\sub{M} - t\msub{\gend} = \gend$, as it should; indeed it happens only for a specific choice of $\gend$, namely, for the case being studied, with $\gend = 6$, an integer value just in view of the one-day resolution of the data. A convenient double check is done on the maximum of $I\subs{sc}(t)$ falling on December 2, 2020 (see fig.~\ref{scaledRemoved}). Very remarkable is the fact that the height of the peaks of $\mathcal{R}_t$ does depend on the value one wants to give to $\gend$, the same way as the days when $\mathcal{R}_{t}$ is equal to $1$ do: so, all of these things are bounded by the SIR model, a fact that must be considered truly important in evaluating the validity of the model. \newline
The estimate $\gend = 6$ days is in total agreement with the average $6.7 \pm 1.9$ days given for Italy in ref.\cite{Cereda et al}: this is a success of eq.~\ref{Rt discrete}\,, that strengthens the agreement, within the uncertainties, of the resulting $\mathcal{R}_{t}$ with that from other algorithms, as those reported in ref.\cite{infn1}\,, with references therein. Fig.~\ref{Fig:Rt} shows this SIR generated $\mathcal{R}_t$, together with the one from eq.~\ref{simple Rt}\,; for either, error bars corresponding to a $\pm 2$ days uncertainty on both $\gend$ and $\remd$\, are also shown.

\section{The ``true'' numbers}

\subsection{The ``corrected'' cumulative and daily-new infections relatively to the swab-confirmed infectious people}
To avoid confusion, it is worth remarking that {\em infections at day t} is meant as the cumulative number of infections up to and including that day, while the number of {\em infectious people} at some day t refers to those people who were infected possibly earlier and are still able to transmit infection at that day. Thus, the daily new {\em infections curve} is different from the {\em infectious curve}.
\newline
Since $N$ in eq.~\ref{N const} is conserved, eq.~\ref{fond.eqs.1} can be written as
\begin{equation}
 \frac{d (N - S)}{dt}(t+\gend) \;=\; \gamma\,\mathcal{R}_t\; I(t)\,, \label{total new infections}
\end{equation}
expressing the daily new infections, gross of {\em removed} people (while the {\em infectious} numbers are {\em net} of {\em removals}). Indeed
\begin{equation}
  \mathcal{T}(t) \,=\, N - S(t) \,=\, I(t) + R(t) \label{total cases}
\end{equation}
is nothing but the total cases of infections at time $t$\,.
\begin{figure}[ht!]
  \centering.
  \includegraphics[width=0.90\textwidth]{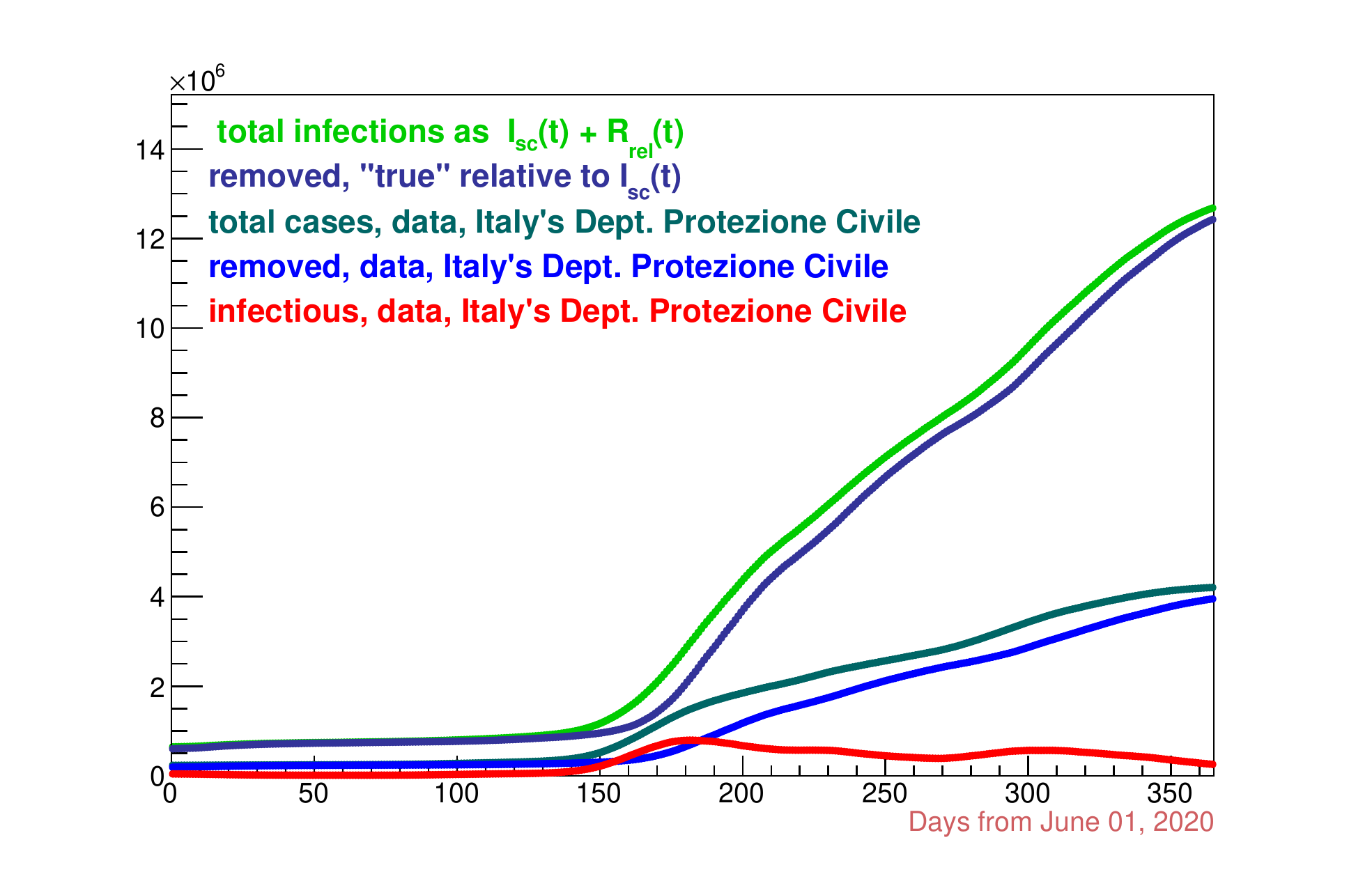}
  \caption {\footnotesize Italy, Covid-19 second through the third waves, estimates of {\em removal} 
    (dark blue curve) and total (green curve) cases as from correction relative to the
    swab-confirmed infectious (red) curve. Also shown data in dark green (total cases) and
    blue ({\em removals})}
  \label{Fig:totalNewSIR}
\end{figure}
For what has been done so far, eq.~\ref{total new infections} must be replaced by 
\begin{subequations}   \label{final relative}
  \begin{align}
    &\frac{d\,\mathcal{T}\sub{rel}}{d t}(t) \;=\; \gamma\,\mathcal{R}_t\; I\sub{sc}(t)\,,
    \label{rel new infections}\\
    &\mathcal{T}\sub{rel}(t) \,=\, I\sub{sc}(t) + R\sub{rel}(t)\,,
    \label{rel total cases}
  \end{align}
\end{subequations}
giving the {\em corrected} cumulative number of infections relatively to the swab-confirmed infectious people only. Fig.~\ref{Fig:totalNewSIR} illustrates eq.s~\ref{final relative}\,.

\subsection{Infections from a-symptomatic and symptomatic infectious people: estimate of the ``true'' numbers}
There are several studies on the relevance of SARS-CoV-2 transmission from asymptomatic people, like \cite{AsymptInfected}, \cite{HiddenChall}, \cite{MagnAsympt}, \cite{infectiousness} and references therein. Quite recent and complete is ref. \cite{infectiousness}, where a decision analytical model is used to assess the proportion of SARS-CoV-2 transmissions in the community, likely occurring from subjects who did not develop any symptom. In that work data from a meta-analysis was used to set the {\em generation time} at a median of 5 days and {\em infectious period} at 10 days, in good agreement respectively with the 6 and 10 days stated in the present work. The reported conclusion is that, across a range of plausible scenarios, a $59\%$ of infection transmission occurs from persons without symptoms: no clear uncertainty is given, but the statement that the figure should be {\em at least} $50\%$, suggests an uncertainty of $\pm 10\,\%$. Also it is stated that the infected individuals who never develop symptoms are $75\%$ as infectious as those who do develop symptoms. 
\begin{figure}[ht!]
  \centering.
  \includegraphics[width=0.95\textwidth]{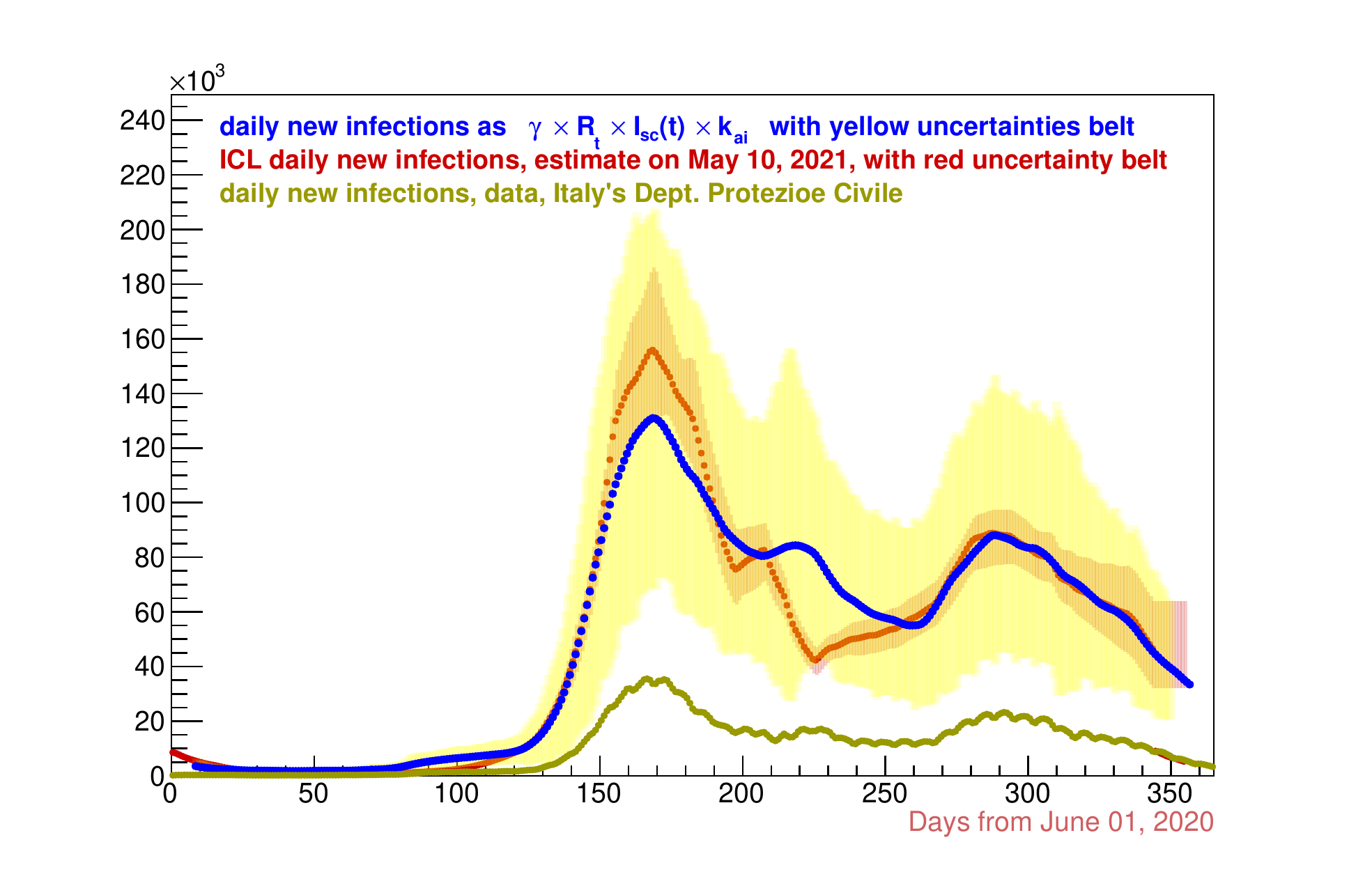}
  \caption {\footnotesize Italy, Covid-19 second through the third waves. Estimates of daily new
    infections this work in blue and uncertainty belt in yellow; ICL's in dark red.}
  \label{Fig:dailyNewSIR}
\end{figure}
\newline
Let's call $f\sub{(asy)}$ the percentage fraction of the asymptomatic infectious subjects over all the infectious people and $i\sub{(asy)}$ their relative infectiousness, that is the percentage fraction of the infectiousness of those who had developed symptoms: then, according to the best available figures, it would be $f\sub{(asy)} = 59 \pm 10\,\%$ and $i\sub{(asy)} = 75\%$, the latter with a trial, very conservative, uncertainty of $\pm 20\%$. \newline
Denoting by $I\sub{(asy)}(t)$ the number of the asymptomatic infectious individuals and recalling that $I\sub{sc}(t)$ indicates the number of the the swab-confirmed infectious subjects, it should be 
\begin{equation}
  I(t) - I\sub{(sc)}(t) \quad=\quad I\sub{(asy)}(t) \quad=\quad
  \frac{f\sub{(asy)}}{100}\,\frac{i\sub{(asy)}}{100}\,I(t)\,,
  \nonumber
\end{equation}
whence
\begin{subequations}
  \begin{align}
    &I(t)  \,=\,
    \left(1 \,+\,\frac{f\sub{(asy)}}{100}\,\frac{i\sub{(asy)}}{100}\right)\,I\sub{(sc)}(t)
    \;\eqdef\; k\sub{(ai)}\,I\sub{(sc)}(t) \,, \label{to_abs} \\
    &R(t) \;=\; k\sub{(ai)}\,R\sub{rel}(t)\,, \label{Rtrue} \\
    &k\sub{(ai)} \;\simeq\; 1 \,+\,\frac{f\sub{(asy)}}{100}\,\frac{i\sub{(asy)}}{100}
    \;\simeq\; 1.44 \,\pm\, 0.22 \,. \label{kai}
  \end{align}
\end{subequations}
Then, in view of eq.~\ref{total cases},\,  eq.~\ref{total new infections} becomes
\begin{equation}
  \frac{d\,\mathcal{T}}{dt}(t+\gend) \;=\; \gamma\;\mathcal{R}_t\; k\sub{ai}\, I\sub{sc}(t)\,,
  \label{totalNewInf}
\end{equation}
with $\mathcal{T}(t)$ the ``true'' cumulative number of infections at day $t$, while its derivative represents the ``true'' daily new infections.\newline
Fig.~\ref{Fig:dailyNewSIR} shows the daily new infections curve, compared with the Imperial College's (ICL) model estimate, as published in \cite{ourworldindataICL}. The model in question is a stochastic SEIR variant, that adopts multiple infectious states, which in turn reflect different COVID-19 severities.
It uses an estimate  of the {\em infectious fatality rate} (IFR), assuming that the number of
confirmed deaths from Covid-19 is equal to the {\em true} Covid-19 deaths number; it also uses an estimate of the {\em effective reproduction number}, based on the changes of the virus transmission rate caused by the average mobility trends. \newline
So, the ICL model's approach is totally different from the one followed in the present work; nevertheless the respective ``true'' daily new infections estimates appear to be in quite good agreement, except on a time interval around 1 January 2021 (day 220 in plots of this paper), where the ICL curve shows a deep local minimum instead of a local maximum as in the data of Italy's Department of Protezione Civile. The uncertainty belt of the ICL estimates are surprisingly narrow.
\begin{figure}[ht!]
  \centering.
  \includegraphics[width=0.90\textwidth]{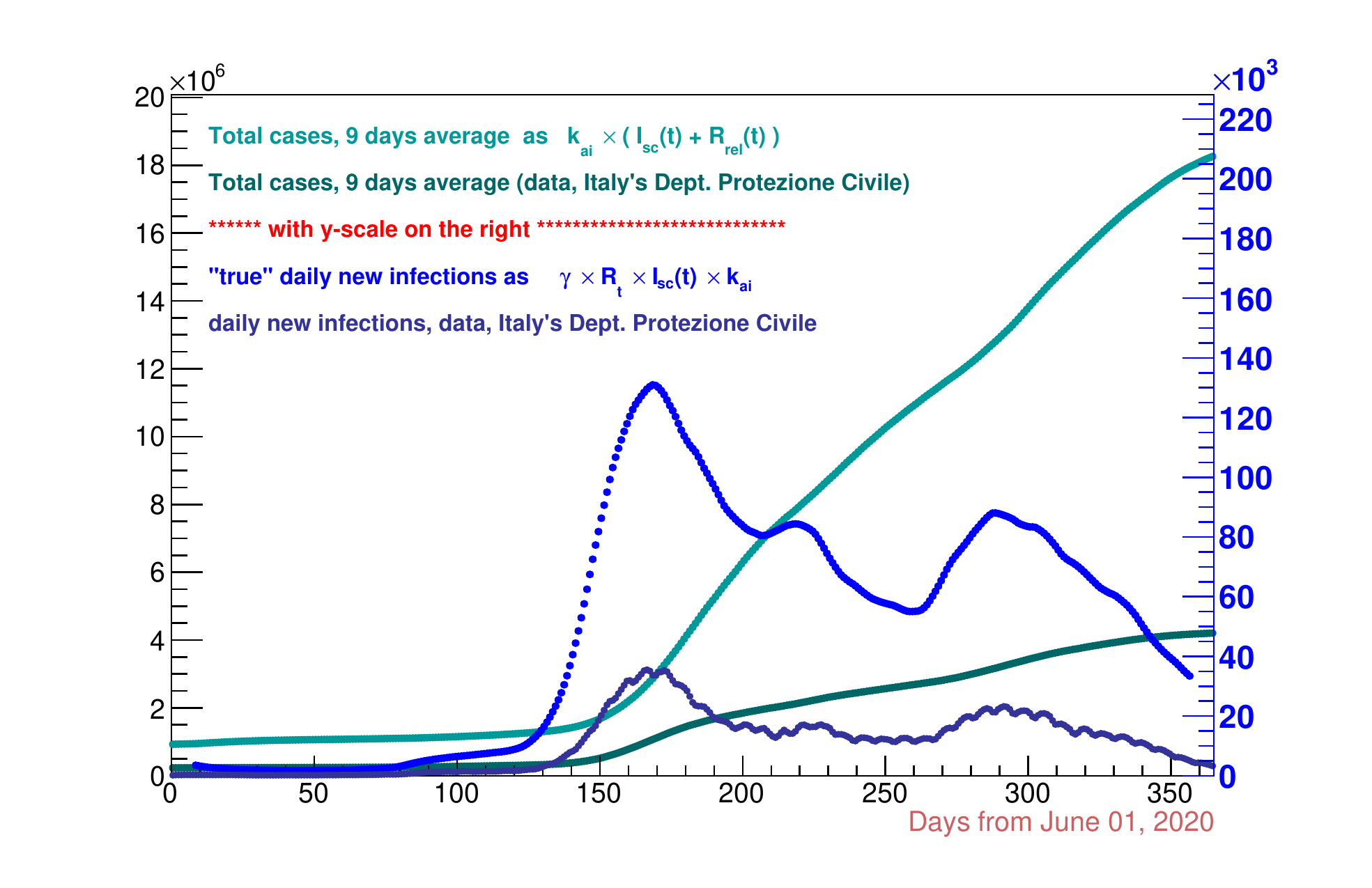}
  \caption {\footnotesize This work estimates of the ``true'' total cases (light green) and
    correspondent data form Italy's Department of Protezione Civile (dark green) with scale on the
    left; this work ``true'' daily new infections (blue) and corresponding official data (dark blue)
    with scale on the right.}
  \label{Fig:total_true}
\end{figure}
In fig.~\ref{Fig:total_true} the present work's estimates of the ``true'' total cases of infections are shown, together with the estimated ``true'' daily new infections, the latter with their own scale on the right; also, the data as from Italy's Department of Protezione Civile are plotted. 
\newline
Incidentally, the ripple visible in fig.~\ref{Fig:dailyNewSIR} and fig.~\ref{Fig:total_true} on the the data from Italy's Department of Civil Protection, has the typical 7-days periodicity that arises from the weekend reduced data recording. 

\section{Conclusions}
Taking as case study the second to the third waves of SARS-CoV-2 in Italy, the SIR model is confronted with data, after reformulating its equations by the explicit introduction of the important {\em effective reproduction number} $\mathcal{R}_t$, as well as the {\em generation time} and the {\em infectious period}, usually, erroneously, neglected. The relationships it sets among the main observables are actually found in the data, in particular between the curve of the swab-confirmed infectious individuals and the curve of the {\em removed} (healed or deceased) subjects. Indeed, taking advantage of its scale invariance and choosing the curve of the swab-confirmed infectious people as a reference, the model suggests a correction on the number of {\em removed} individuals for just a factor which would take into account: a) infected people who have not developed relevant symptoms and, therefore, were not detected; b) deaths erroneously not attributed to Covid-19. {\em Generation time}, {\em infectious period} and {\em effective reproduction number} have been sought from the data through the model. At the very end, the curve of the swab-confirmed infectious individuals has been completed for the proportion of infection transmissions likely occurred from individuals with no symptoms, using figures published in important works ( \cite{AsymptInfected}, \cite{HiddenChall}, \cite{MagnAsympt}, \cite{infectiousness} ). Thus an estimate of the {\em ``true numbers''} of the pandemic in Italy is obtained for the considered period of time. All the results are in good agreement with those of other studies, in particular of the ICL group (\cite{IC}), whose approach is totally different from the present. The vision on and use of the SIR model of this work are new.

% 

%\newpage

%\bibliography{SIRDbibfile}

%\begin{thebibliography}{}
%\input{SIRD.bbl}
%\end{thebibliography}

\end{document}